\documentclass[twocolumn]{article}

\usepackage{parts/_preamble}
\newacronym{AI}{AI}{artificial intelligence}
\newacronym{ML}{ML}{machine learning}

\newacronym{ANN}{ANN}{artificial neural network}
\newacronym{SNN}{SNN}{spiking neural network}
\newacronym{RNN}{RNN}{recurrent neural network}
\newacronym{D2D}{D2D}{device-to-device}
\newacronym{IoT}{IoT}{Internet of things}
\newacronym{DPE}{DPE}{dot-product engine}
\newacronym{RTN}{RTN}{random telegraph noise}
\newacronym[longplural=redox-based resistive random-access memories]{ReRAM}{ReRAM}{redox-based resistive random-access memory}
\newacronym[longplural=phase-change memories]{PCM}{PCM}{phase-change memory}
\newacronym{MCU}{MCU}{microcontroller unit}
\newacronym{FPGA}{FPGA}{field-programmable gate array}
\newacronym{DRAM}{DRAM}{dynamic random-access memory}

\newacronym{vdW}{vdW}{van-der-Waals}
\newacronym{FET}{FET}{field-effect transistor}
\newacronym{GFET}{GFET}{graphene field-effect transistor}
\newacronym{FinFET}{FinFET}{fin field-effect transistor}
\newacronym{MOSFET}{MOSFET}{metal--oxide--semiconductor field-effect transistor}
\newacronym{NCFET}{NCFET}{negative capacitance field-effect transistor}
\newacronym{TFET}{TFET}{tunnelling field-effect transistor}
\newacronym{MCFET}{MCFET}{multi-channel field-effect transistor}
\newacronym{FeFET}{FeFET}{ferroelectric field-effect transistor}
\newacronym{TMD}{TMD}{transition metal dichalcogenide}
\newacronym{EDP}{EDP}{energy-delay product}
\newacronym{BTBT}{BTBT}{band-to-band tunnelling}
\newacronym{SS}{SS}{subthreshold swing}
\newacronym{QD}{QD}{quantum dot}
\newacronym{JJ}{JJ}{Josephson junction}
\newacronym{SPE}{SPE}{single-photon emitter}
\newacronym{BLG}{BLG}{bilayer graphene}
\newacronym{GAAFET}{GAAFET}{gate-all-around field-effect transistor}
\newacronym{SMC}{SMC}{semimetal chalcogenide}
\newacronym{MIM}{MIM}{metal--insulator--metal}
\newacronym{hBN}{hBN}{hexagonal boron nitride}

\newacronym[
    prefixfirst={a\ },
    prefix={an\ }
]
{MTJ}{MTJ}{magnetic tunnel junction}
\newacronym{TMR}{TMR}{tunnel magnetoresistance}
\newacronym{GMR}{GMR}{giant magnetoresistance}
\newacronym{STT}{STT}{spin transfer torque}
\newacronym{SOT}{SOT}{spin-orbit torque}
\newacronym{VCMA}{VCMA}{voltage-controlled magnetic anisoropy}
\newacronym{MRAM}{MRAM}{magnetoresistive random-access memory}
\newacronym{SRAM}{SRAM}{static random-access memory}
\newacronym{PIM}{PIM}{processing-in-memory}
\newacronym{LIM}{LIM}{logic-in-memory}
\newacronym{CMOS}{CMOS}{complementary metal--oxide--semiconductor}
\newacronym{STNO}{STNO}{spin-torque nano-oscillator}
\newacronym{CVD}{CVD}{chemical vapor deposition}
\newacronym{ASL}{ASL}{all spin logic}

\title{\titleMeta}

\begin{document}

\newgeometry{onecolumn, margin=30mm}

\date{}

\maketitle

\begin{abstract}
    In a data-driven economy, virtually all industries benefit from advances in information technology---powerful computing systems are critically important for rapid technological progress.
However, this progress might be at risk of slowing down if we do not address the discrepancy between our current computing power demands and what the existing technologies can offer.
Key limitations to improving energy efficiency are the excessive growth of data transfer costs associated with the von~Neumann architecture and the fundamental limits of \glsentryfull{CMOS} technologies, such as transistors.
In this perspective article, we discuss three technologies that will likely play an essential role in future computing systems: memristive electronics, spintronics, and electronics based on 2D~materials.
We present how these may transform conventional digital computers and contribute to the adoption of new paradigms, like neuromorphic computing.

\end{abstract}

\thispagestyle{empty}

\restoregeometry

\section{Introduction}

Computers have become an integral part of the modern world.
Technologies from instant messaging to searches on the Internet to smart assistants are enabled by devices that perform logical operations and store information over time.
With such an explosion of uses, it is not surprising that energy costs have been increasing too---some estimate that information and communications technology could constitute from \SI{8}{\percent} to \SI{21}{\percent} of the global electricity demand by the end of the decade~\cite{Jo2018}.
Of course, some applications may contribute to this more than others.

Most notably, \gls{AI} and \gls{ML} have become indispensable in a wide range of rapidly growing data-centric technologies, including the \gls{IoT}, transport, medicine, security, and entertainment.
It is now recognized that \gls{AI} might have a hardware problem~\cite{Na2018} associated with huge computational demands, which are directly reflected in the energy consumption.
This is not sustainable and is rapidly becoming a critical societal challenge.
The soaring demand for computing power in \gls{ML} vastly outpaces improvements made through Moore's scaling or innovative architectural solutions.
From 2012 to 2020, hardware performance of state-of-the-art \gls{AI} has improved by a factor of \num{317}~\cite{MeKe2022}; this is not enough to meet the growing computing demands of \gls{AI} applications.
The size of state-of-the-art \gls{AI} models has been increasing exponentially, as have their training costs---from a few dollars in 2012 to millions of dollars in 2020~\cite{Ar2021}.
A pressing need to develop novel technologies to address this issue at the fundamental level and build efficient \gls{AI} systems has recently become acute.
More fundamentally, there is a great need for low-energy computing elements, including those based on different physical principles than \gls{CMOS} transistors implementing Boolean logic.

This perspective article will discuss memristors, spintronics, and 2D~materials and devices, explaining how they can both improve current computing hardware and enable new computing paradigms.
We will present the main physical principles and the promise of these technologies, as well as some materials and engineering challenges that must be addressed before full adoption.
The role of these emerging technologies will be discussed both in the context of conventional computing, which is based on digital electronics and Boolean algebra, and promising new approaches like neuromorphic computing.
This is by no means an exhaustive review and does not imply that other technologies and approaches are not going to play an important role; many alternatives will likely complement the systems we discuss here.
Furthermore, the three approaches we present often overlap---at the extreme, we might even have spintronic memristors partially based on 2D~materials~\cite{ShWa2021}.

\subsection{Basic principles}

\subsubsection{Memristors}

Memristor was formalized as a circuit element in 1971~\cite{Ch1971}---an electrical property, called memristance, relating electric charge and magnetic flux was introduced.
Memristor's existence was motivated by the fact that this relation filled a gap in fundamental symmetries observed in circuit theory.
Since late 2000s, there has been a rebirth of interest in memristors, followed by various physical implementations.
The landscape of memristive technologies and the underpinning physical mechanisms is vast and still rapidly expanding~\cite{IeWa2015}.

Memristors, in most cases, are based on the concept of resistance switching.
Resistance switching is a reversible process where a memristor changes its resistance with externally applied electrical stimuli.
In most cases, resistance switching results in nonvolatile states with long retention times even after the stimuli are removed---the memristive device ``memorizes'' the resistance state.
However, resistance switching can also be achieved by other types of stimuli (e.g.\ optical) and could lead to volatile switching, which benefits particular applications (e.g.\ neuronal spiking).

\begin{figure*}[h!]
  \centering
  \includegraphics{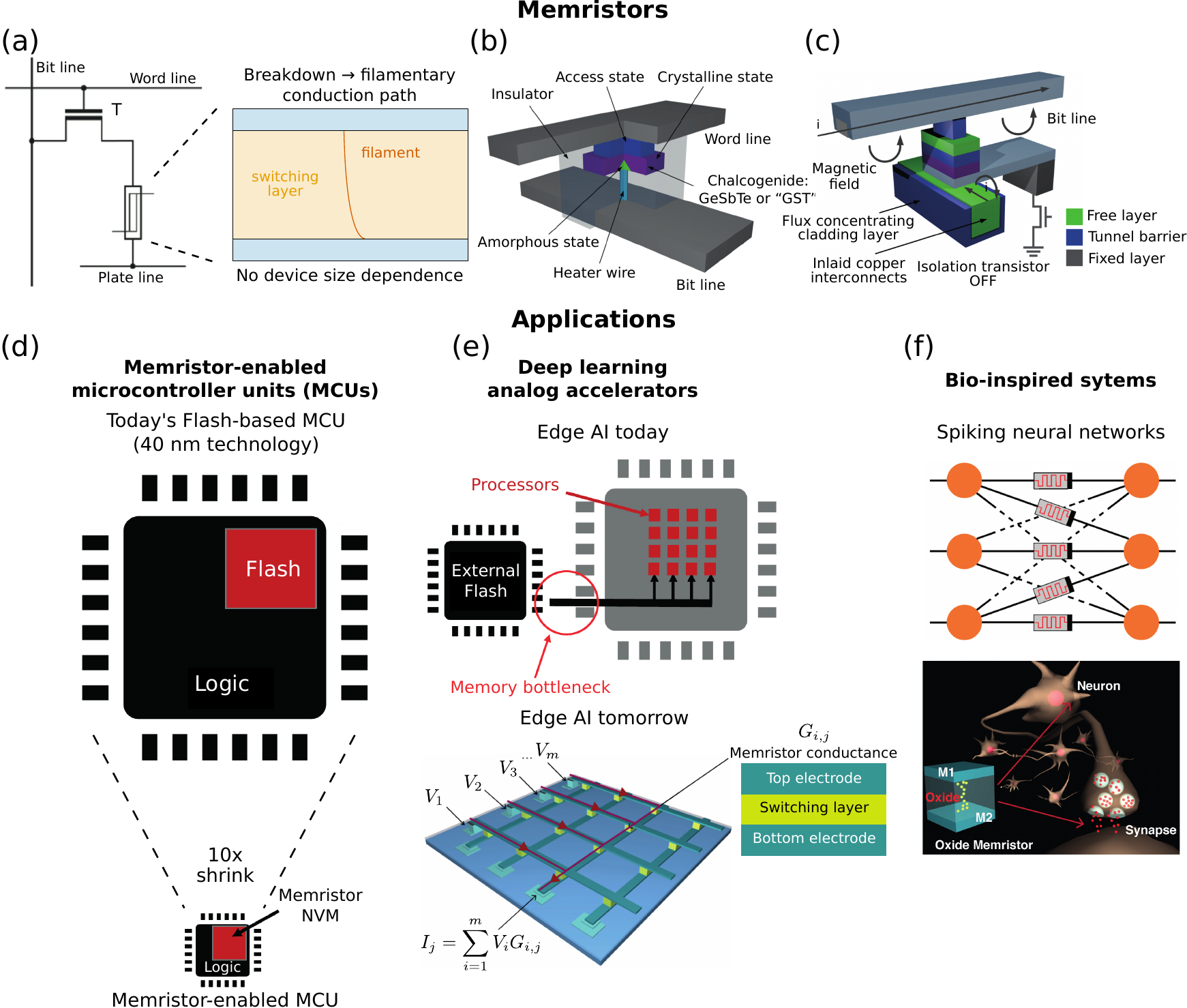}
  {\phantomsubcaption\label{fig:memristors:a}}
  {\phantomsubcaption\label{fig:memristors:b}}
  {\phantomsubcaption\label{fig:memristors:c}}
  {\phantomsubcaption\label{fig:memristors:d}}
  {\phantomsubcaption\label{fig:memristors:e}}
  {\phantomsubcaption\label{fig:memristors:f}}
  \mycaption{%
    Overview of memristive devices and their potentials uses in computing%
  }{%
    There exist multiple memristive technologies, including (a)~\glsentrylong{ReRAM}, (b)~\glsentrylong{PCM}, and (c)~\glsentrylong{MRAM}.
    Common applications of memristive devices include (d)~embedded non-volatile memory, (e)~analog deep learning accelerators based on programmable crossbars, and (f)~bio-inspired systems implemented by memristors that emulate synapses and neurons.
  }\label{fig:memristors}
\end{figure*}

There exist many memristive technologies, but most rely on similar physical principles.
Three examples of such technologies---\gls{ReRAM}, \gls{PCM}, and \gls{MRAM}---are shown in \cref{fig:memristors:a,fig:memristors:b,fig:memristors:c}.
Memristors are typically implemented as simple two-terminal capacitor-like structures, where a switching layer is sandwiched between two electrodes.
The resistance of the switching layer can be programmed to various resistance states with the application of voltage pulses.

Resistance switching manifests itself slightly differently in \glspl{ReRAM}, \glspl{PCM}, and \glspl{MRAM}.
In \gls{ReRAM} technologies, resistance switching is based on the creation/dissolution of conductive filaments (intrinsic to the oxide layer or a result of metallic diffusion from electrodes); local nanoionic redox phenomena drive resistance switching in \glspl{ReRAM}.
There are different flavors of \gls{ReRAM} devices, but they can be broadly divided by the type of switching: (1) intrinsic switching, which manifests itself as an intrinsic property of the switching material and (2) extrinsic switching, which is controlled by indiffusion (typically from metal electrodes) and drift of metal ions extrinsic to the fabricated switching layer~\cite{MeKe2015}.
Alternatively, the devices may be classified by the dominant driving forces of the switching process; this would result in electrochemical metallization cells, valence change \glspl{ReRAM}, and thermochemical \glspl{ReRAM}~\cite{IeWa2015}.
In \glspl{PCM}, the switching is governed by the reversible process of crystallization and amorphization of phase-change materials.
Finally, the programmable relative spin orientation of two ferromagnetic layers is the basis of \gls{MRAM} operation.

It is important to note that novel devices based on different resistance switching mechanisms are still being developed.
Notable examples include nanometallic memristors~\cite{LuAl2019}, which rely purely on electronic effects, and \ce{Ti}/\ce{ZnO}/\ce{Pt} structures that rely on carrier trapping/detrapping of the trap sites~\cite{PaLi2016}.
Such devices could provide further improvements in terms of speed, uniformity, and low-power operation.

As shown in \cref{fig:memristors:d,fig:memristors:e,fig:memristors:f}, a wide range of memristor applications have been suggested, including embedded digital non-volatile memory, analog deep learning accelerators, and neuromorphic spiking systems~\cite{MeSe2020}.
We discuss these and other potential applications in more detail later in the text.
We suggest consulting rich literature for details and descriptions of different physical mechanisms and many more types of memristive devices and technologies~\cite{RaWe2010,GrQu2016,IeWa2015}.

\subsubsection{Spintronics}

Conventional electronic systems rely on electron charges---these systems use voltage levels and currents to process information.
But the electron has another intrinsic property, called ``spin,'' making it analogous to a tiny magnet.
The core concept of spintronics is to use this degree of freedom to create functional electronic devices that cannot be realized using conventional semiconductor technologies.
Magnets can store digital information cheaply and reliably due to their excellent nonvolatile property; combining this with spin-dependent transport for efficient writing and read-out is a viable approach to making disruptive innovations in the electronic device market.

The quantum mechanical Pauli exclusion principle and the Coulomb interaction generate the so-called exchange coupling between spins, creating the magnetic orders of spin ensembles, with the order parameter of magnetization $\matr{M}$.
The central concept of spintronics is to store information bits in local $\matr{M}$ that can be electrically written and read in an energy-efficient manner for data storage and processing~\cite{HIROHATA_JMMM2020,Zutic_RMP2004,Wolf_Science2001}.
The magnetic field $\matr{H}$ is a conventional way to control $\matr{M}$ via the Zeeman interaction ($-\matr{M} \cdot \matr{H}$), e.g.\ when the two vectors are aligned in parallel, the free energy of the system becomes lower, hence stabilized.
Magnetic moments are nonvolatile in general, meaning that when we switch off magnetic fields, the size and direction of the moments are unchanged.
This is possible due to the presence of the aforementioned exchange interaction and magnetic anisotropies.

\begin{figure*}[h!]
  \centering
  \includegraphics{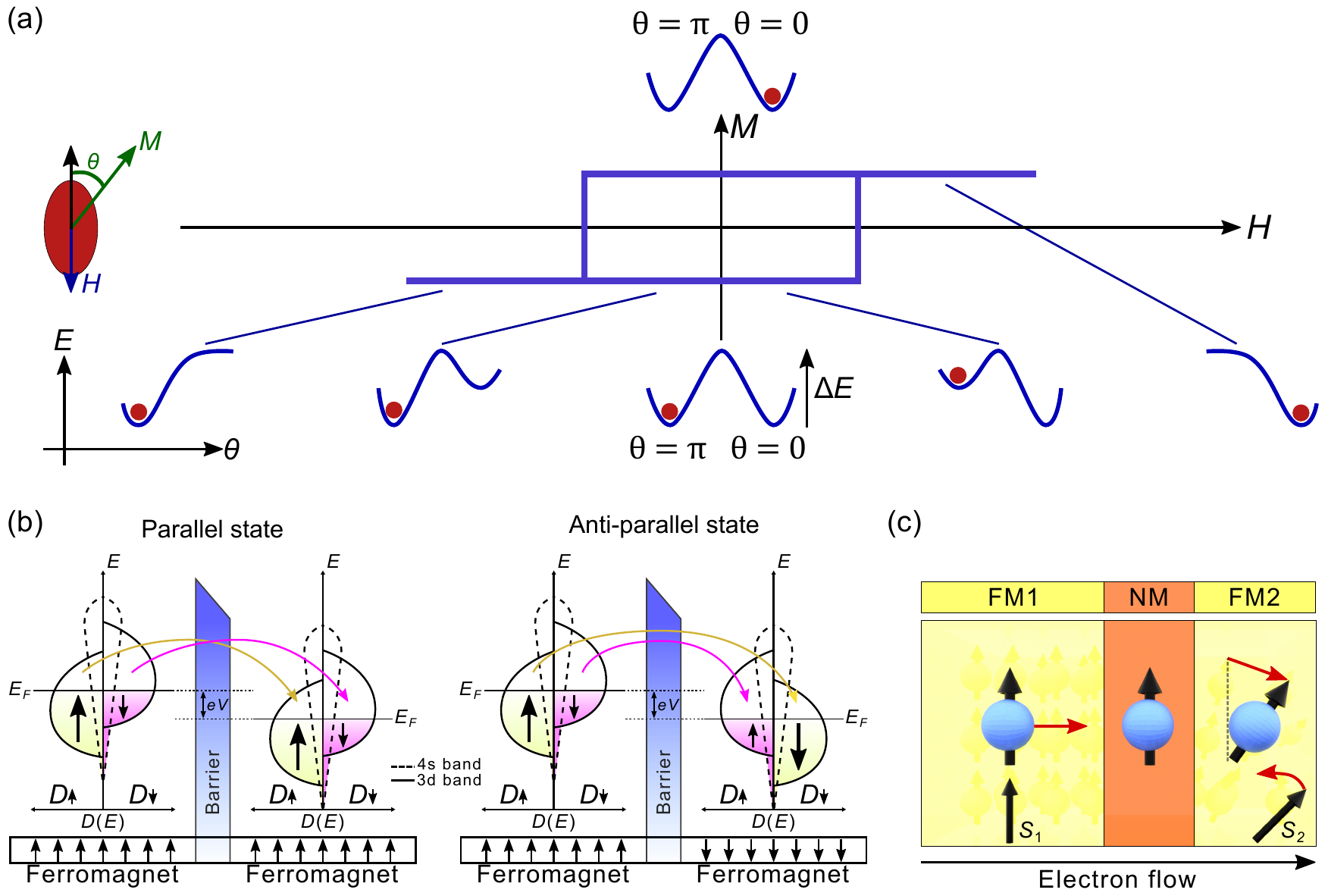}
  {\phantomsubcaption\label{fig:spintronics:a}}
  {\phantomsubcaption\label{fig:spintronics:b}}
  {\phantomsubcaption\label{fig:spintronics:c}}
  \mycaption{%
    Basic principles of spintronics%
  }{%
    (a)~Magnetic switching in a magnet with uniaxial anisotropy.
    When we apply magnetic fields along the easy axis of the uniaxial anisotropy, the magnetic free energy as a function of angle $\theta$ changes.
    At the point where the energy barrier is removed by the Zeeman energy, the magnetization switching occurs as a jump of $M$.
    (b)~Schematic of tunnel magnetoresistance with the density of states $D(E)$ for two magnetic electrodes for \glsentryshort{MTJ}.
    For the parallel configuration (left image), the large $D(E)$ of the up-spin electrons at the Fermi level can produce large tunneling probability proportional to $D_1^2$.
    For the anti-parallel case, the tunneling probability is smaller due to the size of $D_1 \cdot D_2$.
    (c)~Schematic of the spin transfer torque mechanism.
    A spin-polarized electron is generated in FM1 and enters into FM2.
    The polarization angle of the conduction electron is tilted in FM2 as a result of angular momentum transfer into $S_2$.
    This produces a magnetic torque on $S_2$.
  }\label{fig:spintronics}
\end{figure*}

In a ferromagnet, where the exchange interaction aligns individual moments along the same direction, flipping one of the magnetic moments against this direction requires large energy cost thus maintaining the total moments along the favored direction\footnote{There is an excitation state of this magnetically ordered system (called magnons) that can be realized by tilting the moments; however, this results in a slight change of the total moments.}.
The equilibrium direction of $\matr{M}$ is determined by the magnetic free energy where---with zero external magnetic field---the magnetic anisotropy creates local minima as a function of angle, as shown in \cref{fig:spintronics:a}.
The energy barrier between the minima characterizes the thermal stability of the moment orientation, directly relevant to the reliability for storing data in a magnetic cell.
If the barrier height $\Delta E$ is too small, an accidental reversal of the magnetic moment can take place, resulting in a data loss, whereas data retention of ten years is generally guaranteed when $\Delta E/(k_\mathrm{B} T) > 60$ in typical magnets.
This mechanism is the origin of nonvolatility in magnetic materials, and optimizing parameters such as $\Delta E$ (the size of magnetic anisotropy) is one of the major topics in spintronic applications.

Another key ingredient for spintronic devices is that transport parameters (e.g.\ resistivity) can be controlled by $\matr{M}$.
In ferromagnets, the density of states at the Fermi level for up and down spin electrons is different due to the energy splitting by the exchange coupling (see \cref{fig:spintronics:b}).
\Glspl{MTJ} exploit this property as \gls{TMR} by having two magnetic layers with a tunnel barrier (\cref{fig:spintronics:b}), in which the tunneling probability depends on the spin polarization of electrons at the Fermi level for each electrode~\cite{JULLIERE_PR1975,Yuasa_JPhysD2007}.
\Gls{TMR} devices exhibit larger resistance changes than \gls{GMR}~\cite{Baibich_PRL1988,Binasch_PRB1989,Tsymbal_SSP2001}, in particular \gls{TMR} devices with a \ce{MgO} barrier~\cite{Parkin_NMater2004,Yuasa_NMater2004}.
A high \gls{TMR} value is critical for reliability of read-out of spintronic devices using \glspl{MTJ} as well as for reducing the read-out time since it realizes a faster rate of voltage changes during reading. 

Normally we switch $\matr{M}$ by applying $\matr{H}$ greater than magnet's anisotropy field, as shown in \cref{fig:spintronics:a}.
However, this writing method is not scalable with downsizing since $\matr{H}$ produced by an electric current is proportional to the absolute value of the electric current, not current density.
As a scalable magnetization switching mechanism, the concept of \gls{STT} was independently proposed by \citeauthor{SLONCZEWSKI_JMMM1996}~\cite{SLONCZEWSKI_JMMM1996} and \citeauthor{Berger_PRB1996}~\cite{Berger_PRB1996}.
In this scheme (\cref{fig:spintronics:c}), spin-polarized currents injected into a magnetic layer can exert torques via angular momentum transfer between the conduction and localized electrons~\cite{RALPH_JMMM2008}.
An electric current through \pgls{MTJ} can switch magnetization of one layer when the current size is sufficiently large.
The size of this switching current density is directly relevant to the power consumption of spintronic memories, like \gls{MRAM}, which stores and processes digital information by flipping $\matr{M}$ in an array of \glspl{MTJ}.
Furthermore, it is also an important parameter for footprint (density) of spintronic arrays since each \gls{MRAM} cell is powered by a \gls{CMOS} transistor underneath, and this element is so far the limiting factor of downsizing of \gls{MRAM}.
Since high current requires a large \gls{CMOS} transistor, a high-density \gls{MRAM} can be achieved when the writing current is small.
Other emerging magnetization control mechanisms include \glspl{SOT} and \gls{VCMA}, for which readers are invited to read Refs.~\cite{Manchon_RevModPhys2019,Shao_TMAG2021,Matsukura_NNANO2015,Nozaki_review2019} for more details.

\subsubsection{2D materials}

\begin{figure*}[b]
  \centering
  \includegraphics{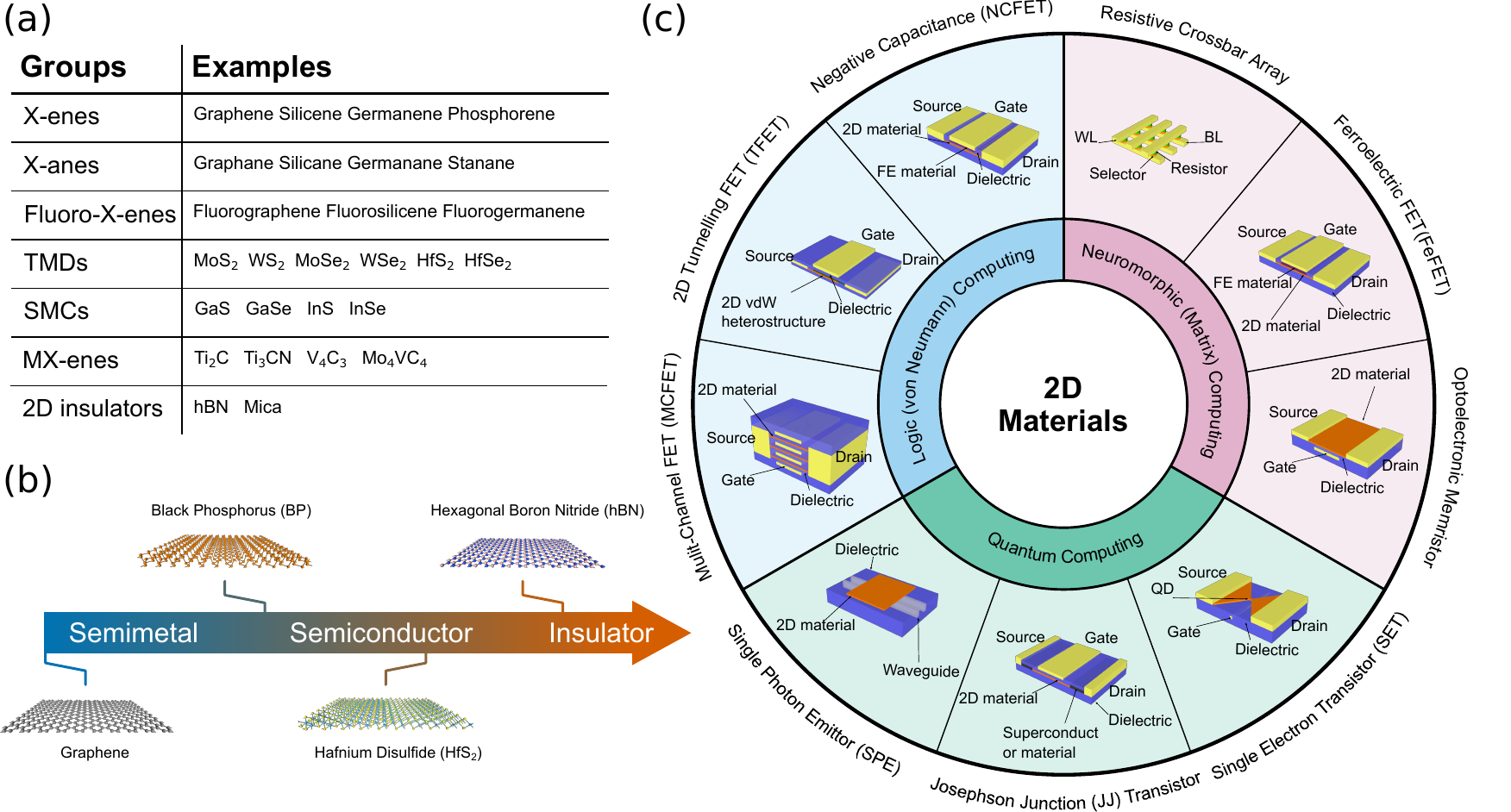}
  {\phantomsubcaption\label{fig:2D:a}}
  {\phantomsubcaption\label{fig:2D:b}}
  {\phantomsubcaption\label{fig:2D:c}}
  \mycaption{%
    Overview of 2D materials and their applications%
  }{%
    (a)~List of the most common 2D materials.
    (b)~The range of 2D materials' electrical properties from zero-bandgap semimetals, such as graphene, to wide-bandgap insulators, such as \glsentrylong{hBN}.
    (c)~Devices based on 2D materials for applications in logic, neuromorphic, and quantum computing.
  }\label{fig:2D}
\end{figure*}

Adopting different computation variables (such as spin) and architectures (such as neuromorphic) leads to a demand for novel materials capable of supporting such technologies.
In this perspective, we also explore two-dimensional layered materials, often simply referred to as two-dimensional~(2D) materials.
We believe that these materials are among the most promising candidates for future computing due to large variety of properties they offer, the possibility of being easily combined into functional structures, and the ease of integration with existing semiconductors and fabrication lines.
2D materials are a large class of materials consisting of stacks of individual layers held together by, typically, van der Waals forces.
Each layer is formed by covalently-bonded atoms and exhibits fully saturated surface bonds, resulting in crystals that are stable even in the form of a single layer, hence the name ``2D'' materials.

\Cref{fig:2D:a} shows a list of the most common 2D materials grouped according to their structure.
X-enes are materials consisting of a single element such as graphene and silicene, whereas X-anes and fluoro-X-enes are their chemical derivatives, e.g.\ graphane refers to hydrogenated graphene and fluorographene---to fluorinated graphene.
\Glspl{TMD} are a class of compounds formed by a transition metal element (\ce{M}) mainly from IV, V or VI group and a chalcogen (\ce{X}), with a generalized formula \ce{MX2}~\cite{2D-11}.
These materials form layered structures of the form \ce{XMX}, with the chalcogen atoms in two hexagonal planes separated by a plane of metal atoms~\cite{2D-11}.
\Glspl{SMC} are similar to \glspl{TMD}; they are formed by a semimetal and a chalcogen, usually occurring in \ce{M2X2} stoichiometry.
MX-enes are ternary layered materials having occurring in the formula \ce{M_{n+1}AX_n} where \ce{M} is an early transition metal, \ce{A} is an element from group $13$ or $14$, \ce{X} is either carbon or nitrogen, and $n$ is an integer between $1$ and $3$.
Finally, the 2D ``library'' also includes insulators, such as \gls{hBN}, an isomorph of graphene consisting of boron and nitrogen atoms.

Despite sharing a similar structure, the properties of 2D materials are incredibly diverse---the ``family'' of 2D materials includes semimetals, direct and indirect bandgap semiconductors, insulators, metals, superconductors, topological, and ferromagnetic insulators, as illustrated in \cref{fig:2D:b}.
The lack of dangling bonds on the surface enables deterministic stacking of different 2D materials to form heterostructure without lattice matching constrains, usually referred as \gls{vdW} heterostructures~\cite{2D-2}.
Such structures have atomically precise control of the thicknesses of the different layers with abrupt interfaces, leading to an unprecedented flexibility in terms of materials and properties available.
Moreover, by controlling the angle between the layers, it is possible to define a Moir\'e superlattice which provides a further degree of freedom, leading to new phenomena (such a superconductivity in twisted bilayer graphene~\cite{2D-3}) and enabling a novel approach to electronics referred to as ``twistronics''~\cite{2D-4}. 

With tens of materials experimentally available and over \num{2000} theoretically predicted~\cite{2D-1}, 2D materials represent one of the most promising material systems for future computing.
From a manufacturing point of view, 2D materials also have significant advantages.
Indeed, these materials are (sub)nanoscopic only in terms of thickness, whereas their lateral dimensions can be macroscopic, leading to a significant technological advantage over other nanomaterials because they can be processed using ``conventional'' semiconductor planar technology~\cite{2D-5}.
Combined with the ease of transferring them from one substrate to another, 2D materials can be easily integrated with existing technologies, particularly at the back-end of line in \gls{CMOS} production lines~\cite{2D-6}.
2D materials are strong candidates for present and future computing paradigms, including logic and neuromorphic computing, as shown in \cref{fig:2D:c}.
Despite being beyond the scope of this perspective, it is worth noting that 2D materials, including \gls{BLG} \gls{QD}~\cite{2D-41}, Josephson junctions~\cite{2D-42} and \gls{hBN} \glspl{SPE}~\cite{2D-43,2D-44,2D-45}, have also been used in the field of quantum computing.
Nevertheless, applications of 2D materials in the field of electronic devices goes beyond what is shown in \cref{fig:2D:c}.
Here, we will provide a prospective overview on how 2D materials can be used as an enabling platform for the technologies discussed.
The reader is invited to read Refs.~\cite{2D-57,2D-58,2D-59,2D-60} for in-depth reviews on the recent progress in the field of 2D electronics.

\section{Conventional Computing Hardware}

Digital computers are the basis of our information and communication technologies.
Logic gates, such as NAND or NOR, implement Boolean algebra, which is used for all digital information processing.
\Glspl{FET}, fundamental building blocks of digital circuits, have followed Moore's scaling law for more than 50 years.
We are still managing to scale transistors; however, the scaling rate has slowed down over the last years~\cite{Wa2016}.
There is a tremendous motivation to investigate post-\glsentryshort{CMOS} technologies, starting from innovations in and understanding of materials and basic nanoscale devices.
\Gls{ReRAM}, spintronic and 2D-based devices could \emph{all} potentially offer better scaling prospects, as well as improved energy efficiency and speed.
These emerging technologies could serve as improved realizations of digital memory and logic, which are used in all conventional, general-purpose computers.

\subsection{Memory}

\Gls{ReRAM}, \gls{PCM}, \gls{MRAM} devices can all be operated as binary memory with two well-defined nonvolatile memory states.
Both \gls{ReRAM} and \gls{MRAM} devices compare favorably against currently used Flash technology, beating it in most performance metrics~\cite{BhSb2017,ZaZu2020}.
\Glspl{MCU} are the first and most attractive applications for these emerging nonvolatile memory technologies.
Today's \glspl{MCU} use embedded NOR Flash, which cannot be easily scaled beyond \SI{28}{\nano\meter} node size; this represents a critical bottleneck, especially considering that more applications are becoming data-intensive (e.g.\ automotive \gls{MCU} needs to operate on a significant amount of data collected by numerous sensors found in modern cars).
Both \gls{ReRAM} and \gls{MRAM} present an attractive opportunity to replace NOR Flash in embedded memory applications offering better scaling (down to most aggressive nodes, \SI{<10}{\nano\meter}) and faster programming/reading speeds (\SI{<5}{\nano\second}).
Beyond embedded memory, \gls{ReRAM} and \gls{MRAM} are also considered as data storage, and thus as a replacement for NAND Flash.
NAND Flash is scalable to most aggressive nodes; however, \gls{ReRAM} and \gls{MRAM} offer better reading speed and lower energy.
Another attractive potential application could replace or augment \gls{SRAM} in edge \gls{AI} applications~\cite{SaPe2018,RiPo2021}, where \gls{ReRAM}/\gls{MRAM} offer similar reading speeds but better scalability and energy efficiency.

In general, \gls{ReRAM}---when used as nonvolatile digital memory---offers
\begin{itemize}
  \item excellent scalability (e.g.\ $10 \times 10$ \si{\nano\meter}~\cite{GoKa2011} and likely below~\cite{ZhMa2011}), which is highly competitive with current memory technologies, like \gls{SRAM} and Flash
  \item large resistance ratio (\num{>10} and much more) critical for fast sensing and reading speeds
  \item fast programming (typically \SI{<100}{\nano\second}, although there are reports of \SI{100}{\pico\second} programming~\cite{AnJo2011})
  \item excellent endurance ($10^{12}$ switching cycles have been reported~\cite{LeLe2011})
  \item small operational energy (e.g.\ sub \si{pJ/bit}~\cite{MiSt2011})
\end{itemize}

\begin{figure*}[b]
  \centering
  \includegraphics{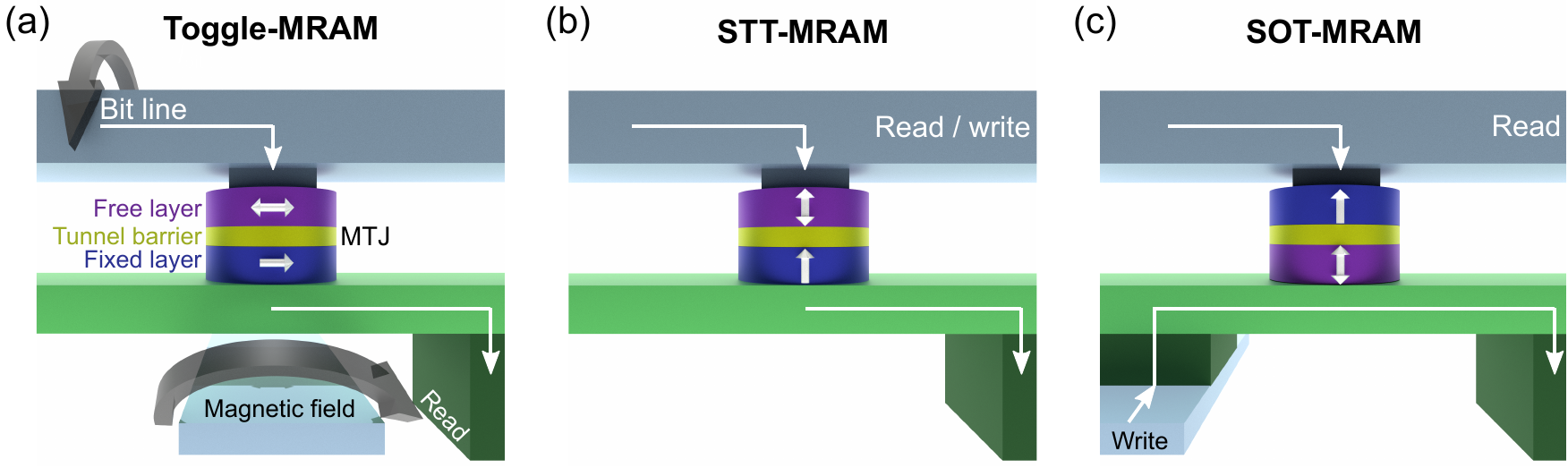}
  {\phantomsubcaption\label{fig:MRAM:a}}
  {\phantomsubcaption\label{fig:MRAM:b}}
  {\phantomsubcaption\label{fig:MRAM:c}}
  \mycaption{%
    Schematics of different MRAM architectures%
  }{%
    (a)~Toggle-MRAM uses magnetic fields to switch magnetization in an MTJ\@.
    (b)~STT-MRAM directly passes an electric current through an MTJ to write their cells.
    (c)~In SOT-MRAM, an electric current flows through the write line, which generates magnetic torques on the layer above.
  }\label{fig:MRAM}
\end{figure*}

In terms of commercialization of \gls{ReRAM}, in 2013, Panasonic released the first \gls{MCU} with embedded \gls{ReRAM}~\cite{Pa2022}.
Many other companies are currently developing \gls{ReRAM} technologies, including Rambus, 4DS, Dialog Semiconductor, Crossbar, Intrinsic Semiconductor Technologies, Weebit Nano, eMemory, and global foundries such as Taiwan Semiconductor Manufacturing Company~(TSMC). 

\Gls{MRAM} consists of an array of \glspl{MTJ} connected with read and write lines for its memory operation.
\Cref{fig:MRAM} displays three different types of individual \gls{MRAM} cells with different writing mechanisms.
In particular, the \gls{STT} writing method has become ripe for industrial applications, and two magnetic layers are magnetized along the perpendicular to the junction plane to minimize footprint.
Non-volatility offers significant advantages in energy saving against volatile memories, such as \gls{DRAM}, which require constant power to maintain their stored information as energy loss.

Major electronics companies have been focusing on \gls{MRAM} development.
Samsung and the partnership between Everspin and Global Foundaries announced their release of a \SI{1}{GB} embedded \gls{MRAM} on their \num{28}/\SI{22}{\nano\meter} technology nodes~\cite{Samsung_2019,GFEverspin_2019}.
The write speed of their technology is orders of magnitude faster than eFlash (\num{200} nanoseconds vs.\ tens of microseconds), with comparable read speeds, providing a power advantage over eFlash in many applications.
Intel announced they are embedding STT-MRAM into devices using its \num{22}-nm FinFET process, with a bit yield rate of greater than \SI{99.9}{\percent}~\cite{Intel_2019}.

STT-MRAMs are believed to be more suitable to embedded memory applications for industrial-grade \glspl{MCU}, autonomous vehicles, and various \gls{IoT} devices~\cite{Guo_IEEE2021}.
Using its high-speed nature, STT-MRAM has been considered as an alternative to \gls{SRAM} applications~\cite{Alzate_IEEE2020} as well as L3/L4 cache replacement, which requires high performance in terms of density, write efficiency, bandwidth, and  endurance~\cite{Sakhare_IEEE2018}.
We point curious readers to more detailed review papers~\cite{Finocchio_JMMM2020,Dieny_NatElec2020} since there is an excellent summary table of STT-MRAM specs against other memory applications.

Emerging writing mechanisms of \gls{MRAM} cells, such as \gls{SOT} and \gls{VCMA}, have been extensively studied for the next generation of \gls{MRAM}~\cite{Guo_IEEE2021,Finocchio_JMMM2020,Dieny_NatElec2020,Shao_TMAG2021}.
Wafer-scale SOT-MRAMs compatible with \gls{CMOS} technologies have been demonstrated~\cite{Garello_VLSIT2019}, together with fast switching demonstration (less than \SI{400}{\pico\second}) in a perpendicularly magnetized SOT-MRAM cell~\cite{Cubukcu_IEEE2018}, show high-speed switching, as well as improved endurance for both standalone-memory and \gls{PIM} applications~\cite{He_IEEE2017}.
\Gls{PIM} refers to performing computational tasks within the memory units where the memory units within these applications need to have high endurance and fast writing/reading since data are more rapidly accessed inside~\cite{Sebastian_NNano2020}.
Combining \gls{SOT} and \gls{STT} writing mechanisms is expected to reduce the writing current down to a range of \num{10}-\num{100}~fJ/bit~\cite{Guo_IEEE2021,Brink_APL2014,Wang_NELEC2018}.

\subsection{Logic}

\subsubsection{\Glsentrylongpl{FET}}

Since the groundbreaking work of Geim and Novoselov that experimentally unveiled the electronic properties of graphene in 2004~\cite{2D-7}, significant attention has been put into its use for transistors.
That is due to graphene's atomic thickness, extremely high room-temperature mobility, saturation velocity and thermal conductivity and the ambipolarity of its field effect.
Because of the lack of bandgap, however, \glspl{GFET} cannot be switched off.
As a result, \glspl{GFET} exhibit only a modest ON/OFF ratio of \num{\sim 10}, which is not suitable for transistor logic applications, where current ratios in excess of $10^4$ are required~\cite{2D-8}.
Nevertheless, \Glspl{GFET} have been used in analog RF electronics, where switching off is not essential, achieving cut-off frequencies in excess of \SI{400}{\giga\hertz}~\cite{2D-9}, and in applications directly benefiting of the ambipolarity of the field effect, such as high-frequency mixers~\cite{2D-10}.

The possibility of isolating individual atomically thin crystals demonstrated by graphene paved the way to the exploration of other 2D materials, in particular \glspl{TMD}.
Molybdenum- and tungsten-based \glspl{TMD}, such as \ce{MoS2} \ce{WS2} and \ce{WS2}, are of particular interest for future transistor logic application as they are atomically thin semiconductors, which can enable reduction of the characteristic length of \glspl{FET} beyond the limit faced by silicon~\cite{2D-12}.
Scaling of body thickness by adopting ultrathin-body on insulator and \gls{FinFET} structures has indeed been key to reduce short-channel effects and extend Moore's law~\cite{2D-13}.
However, the reduction of body thickness in bulk semiconductor below \SI{\sim5}{\nano\meter} is accompanied by a rapid decrease of charge carrier mobility due to thickness variation, dangling bonds and roughness, resulting in a limit to further scaling~\cite{2D-14}.
Conversely, 2D semiconductors have thickness \SI{<1}{\nano\meter} (e.g.\ single layer \ce{MoS2} \SI{\sim 0.65}{\nano\meter}) and mobility in excess of \SI{100}{\centi\meter^2/\volt\second}, significantly higher than sub-\SI{5}{\nano\meter} silicon~\cite{2D-12}.
Moreover, in 3D semiconductor there is usually a tradeoff between bandgap and effective mass and therefore mobility.
Materials with higher bandgap normally show larger effective mass and lower mobility, imposing a compromise between performance and power consumption.
This is not the case in 2D semiconductors, where the mobility is determined by phonon scattering~\cite{2D-15} thus enabling materials combining large bandgap and high mobility.
Saturation velocity also plays a very important role in ultra-scaled devices, where the in-plane field is can easily exceed \SI{1}{\kilo\volt\centi\meter^{-1}}; however, the data available for \glspl{TMD} are scattered and would require a more thorough investigation.
\Glspl{TMD} are extremely interesting candidates for future \gls{MCFET} to reduce the scaling length of \glspl{FET} beyond the limits imposed by silicon.

\subsubsection{\Glsentrylongpl{TFET}}

One of the main figures of merit when assessing \gls{CMOS} efficiency is the energy-delay product of its \glspl{MOSFET}.
One of the main factors governing the \gls{EDP} is the \gls{SS}, which is a measurement of the gate voltage required to change the drain current by a factor of ten.
\Gls{SS} in \glspl{MOSFET}, regardless of the channel material, is thermodynamically limited by the Boltzmann limit.
In \glspl{MOSFET},
\begin{equation*}
  \mathrm{SS} = k_\mathrm{B} T \ln(10) \left( 1 + \frac{C_\mathrm{s}}{C_{\mathrm{ox}}} \right)
\end{equation*}
where $C_\mathrm{s}$ and $C_\mathrm{ox}$ are the semiconductor capacitance (or depletion layer capacitance) and the gate dielectric capacitance, respectively.
It is clear that even if $C_\mathrm{ox} \gg C_\mathrm{s}$, \gls{SS} will never drop below $k_\mathrm{B} T \ln(10)$ ($\approx \SI{60}{\milli\volt/dec}$ at room temperature).

An alternative to thermionic injections over an energy barrier are \glspl{TFET}.
They rely on \gls{BTBT}, resulting in \gls{SS} not limited to \SI{60}{\milli\volt/dec}.
However, to achieve steep \gls{SS} beyond the thermal limit, the energy window for tunneling needs to be sharp, which can only be attained with very abrupt interface.
This has proven to be challenging in conventional planar homojunction \glspl{TFET} because controlling the doping profile to the atomic level is extremely difficult.
Bulk heterojunction \glspl{TFET}, on the other hand, have been demonstrated to outperform their homojunction counterpart.
Nevertheless, the fabrication of such sharp interface is still challenging.

2D materials, owing to their inherently atomically flat surfaces, are well suited for such applications as they can form a sharp interface ideal for tunneling.
Different material combinations have been explored, such as graphene/boron nitride/graphene~\cite{2D-16}, graphene/\ce{WS2}/graphene~\cite{2D-17}, \ce{MoS2}/\ce{WSe2}~\cite{2D-18}, black phosphorus/\ce{SnS22}~\cite{2D-19} and \ce{SnS2}/\ce{WSe2}~\cite{2D-20}.
More interestingly, heterostructures between a 2D materials and a 3D conventional one can bring the best of both worlds.
In particular, \ce{MoS2}/germanium \glspl{TFET} have been reported to achieve ``record'' \gls{SS} of \SI{3.9}{\milli\volt/dec} at room temperature, combined with higher current density compared to other sub-thermionic transistors~\cite{2D-21}.

\subsubsection{\Glsentrylongpl{NCFET}}

Steep \gls{SS} can also be attained by modifying the gating mechanism in \glspl{MOSFET}.
In these devices, the gate controls the channel through direct capacitive approach.
\gls{NCFET} utilizes ferroelectric~(FE) materials, which exhibit metastable spontaneous polarization, which can be triggered through an external field from a low state to high state.
\Glspl{NCFET} employ this abrupt change to switch the device from low (OFF) state to high (ON) state.
However, it is important to note that an appropriate dielectric material (DE) needs to be connected in series with the FE layer to stabilize the negative capacitance state and reduce hysteresis~\cite{2D-22,2D-23}.
The aforementioned \gls{SS} formula needs be changed to include the FE layer effect.
Hence,
\begin{align*}
  \mathrm{SS} &= k_\mathrm{B} T \ln(10) \left(1 + \frac{C_\mathrm{s}}{C_\mathrm{FE} + C_\mathrm{ox}} \right) \\
              &= k_\mathrm{B} T \ln(10) \left( 1 - \frac{C_\mathrm{s}}{|C_\mathrm{FE}| - C_\mathrm{ox}} \right)
\end{align*}
where $C_\mathrm{FE}$ is the capacitance of the FE layer~\cite{2D-24}.

It is clear that to achieve sub-\SI{60}{\milli\volt/dec} \gls{SS}, $C_\mathrm{ox}$ must be larger than $|C_\mathrm{FE}|$, which adds another criterion for choosing the suitable dielectric.
As in \glspl{MOSFET}, \glspl{NCFET} benefits from improved gate control that 2D materials exhibit due to their thinness.
Hence, \gls{SS} as low as \SI{25}{\milli\volt/dec} has been achieved in \ce{MoS2} \gls{NCFET} with \ce{Hf_{0.5}Zr_{0.5}O2} FE with low hysteresis (\SI{\sim 28}{\milli\volt})~\cite{2D-25}.
In addition, based on the industrial direction for \glspl{MOSFET}, we expect that an all-2D-stacked negative-capacitance \gls{GAAFET} that can combine steep \gls{SS} and high ON current would be one of the most promising structures for future logic devices.

\subsubsection{Memristor-based logic}

There are several ways of using memristors for digital logic.
For instance, memristors have been considered as programmable switches for \glspl{FPGA} in the past~\cite{StLi2005,SnWi2007}.
Although, currently, these switches are implemented using \gls{SRAM}, memristor-based switches could lead to significantly improved energy efficiency, e.g.\ reducing cell area by \SI{40}{\percent} and energy-delay-product, by \SI{28}{\percent}~\cite{SaWo2016}.
Alternatively, memristors could be used to implement IMPLY\footnote{Implication $p \implies q$ is false only when $p$ is true and $q$ is false.} logic gates~\cite{BoSn2010}.
The interest comes from the fact that an IMPLY gate with the FALSE operation\footnote{FALSE operation always yields a logical zero.} comprises a complete logic structure.
Memristive implementation of this fundamental logic element could lead to memristor-based logic circuits.
More details and performance comparisons involving this approach can be found in Ref.~\cite{KvSa2013}.

\section{Future Computing Hardware}

While existing compute infrastructure based on Boolean algebra offers many advantages, new hardware paradigms can
\begin{itemize}
  \item improve the efficiency of existing computing tasks
  \item implement functionality that would be infeasible to realize using conventional computers
\end{itemize}
One example is neuromorphic computing, which mimics the structure and/or operation of the brain~\cite{Fu2016}.
Neuromorphic computing can encompass efficient implementations of both well-established concepts, like \glspl{ANN}, and exotic approaches to information processing, like \glspl{SNN} and reservoir computing.
This paradigm aims to perform complex tasks, including recognition and classification, with little energy~\cite{roy_nature2019, chakraborty_APR2020, wei_NeuromorphicDevices2021}.
Multiple emerging technologies hold promise of making these new approaches to computing hardware a reality.

\subsection{\Glsentrylongpl{ANN} on crossbar arrays}

\Glspl{ANN} are implemented on digital computers, but they are very resource-intensive because of (1)~large amounts of data being processed and (2)~the nature of conventional computer architectures.
Modern neural networks can often have billions of parameters~\cite{ShMi2017}, and von~Neumann architecture, which most computers are built around, is not well suited to handle such large models.
Time and energy is mostly spent \emph{not} on performing computations, but on repeatedly moving data between memory and computing units~\cite{ZiSt2018}.

\begin{figure*}[b]
  \centering
  \includegraphics{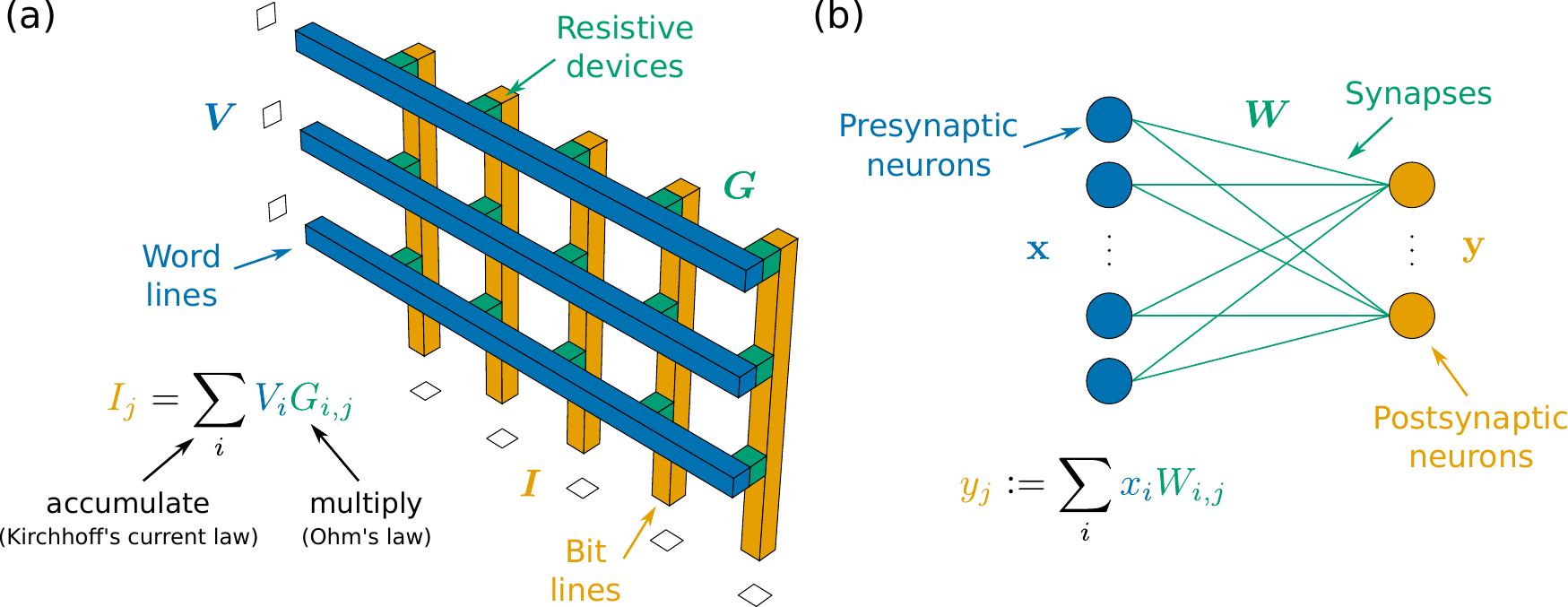}
  {\phantomsubcaption\label{fig:dpe-and-synaptic-layer:a}}
  {\phantomsubcaption\label{fig:dpe-and-synaptic-layer:b}}
  \mycaption{%
    The computing principles behind crossbar-array-based \glsentrylongpl{DPE} and fully connected synaptic layers%
  }{%
    (a)~Using resistive devices in each of the vertical (bit) lines, crossbar arrays can compute dot products of voltages and conductances.
    When multiple of these bit lines are combined, one can compute products of voltage vectors $\matr{V}$ and conductance matrices $\matr{G}$.
    (b)~Synapses in neural networks scale the incoming signals.
    Before nonlinear transformations, these scaled signals are added together by the postsynaptic neurons.
  }\label{fig:dpe-and-synaptic-layer}
\end{figure*}

Resistive crossbars---one of the simplest examples of neuromorphic hardware---may offer a solution to this problem.
In these structures, resistive elements are arranged in an array, as seen in \cref{fig:dpe-and-synaptic-layer:a}.
Ohm's law achieves multiplication of voltages and conductances, while Kirchhoff's current law achieves addition of currents.
With the crossbar structure, these are combined, producing multiply-accumulate operations, or multiplication of voltage vectors and conductance matrices.
By using pairs of devices~\cite{LiHu2018}, the principle can be easily extended to handle negative numbers thus achieving in-memory multiplication of \emph{arbitrary} vectors and matrices.
Such crossbar are usually referred to as \glspl{DPE}.

Hardware acceleration of linear algebra operations is easily applicable to \gls{ML} and \glspl{ANN} in specific.
Fully connected neural networks heavily rely on vector-matrix multiplication to compute outputs of the synaptic layers; this is demonstrated in \cref{fig:dpe-and-synaptic-layer:b}.
During training, optimal weights $\matr{W}$ are determined; this is typically done using gradient descent~\cite{Ru2016}.
After that, during a process called inference, only the inputs $\matr{x}$ change---with each new example, outputs $\matr{y}$ are either used for prediction directly or are passed along to the next synaptic layer.
The fact that weights do not change during inference is one of the primary reasons\footnote{One may also use crossbar arrays to \emph{train} the \glspl{ANN}, as will be explored later.} why crossbars are an appealing candidate for their physical implementation.
Inference can be accelerated by encoding weights into conductances and inputs---into voltages.
The ability of \glspl{DPE} to compute vector-matrix products means that, this way, the synaptic layers of \glspl{ANN} can be implemented in memory, i.e.\ there is no need to transfer the weights during computation, only the inputs have to be applied in the form of voltage vectors.

Easily programmable resistive devices are perfect candidates for \gls{DPE} implementations.
Memristors are one example of such devices---one may encode matrix values into the conductances of memristors embedded in the crossbar array.
Such programming can be done using voltage pulses, which require very little energy~\cite{NiEm2018}.
Examples of such devices include \ce{Ta}/\ce{HfO2}~\cite{LiHu2018} and \ce{SiO_x}~\cite{MeSh2018} memristors.
Spintronic devices can also be used to emulate synaptic behavior---\glspl{MTJ} can act as a local nonvolatile digital memory or as a continuously varying resistance~\cite{Romera_Nature2018, sengupta_roy_2018, song_NatureElect2020}.
For example, the conductance of a three-terminal \gls{MTJ} device can be encoded by controlling the magnitude and the direction of the current flowing through the underlying heavy-metal layer~\cite{sengupta_roy_2018}.

Several neuromorphic proof-of-concept devices have also been realized using 2D materials.
That includes atomically thin \ce{MoS2} memristors having switching ratio ${>} 10^4$ and stable operation up to \SI{50}{\giga\hertz}~\cite{2D-26}, memristors consisting of multilayer \ce{MoS2} encapsulated between graphene layers capable of high temperature (\SI{>300}{\celsius}) operation~\cite{2D-27}, lithium-ion intercalated few-layer metal dichalcogenides and phosphorus trichalcogenides~\cite{2D-29}.
Different switching mechanisms have been identified in 2D materials, including formation of conductive filaments~\cite{2D-30}, grain boundary migration~\cite{2D-31}, phase transition~\cite{2D-32}, oxygen migration~\cite{2D-27}, and graphene has been showed to improve the $I_\mathrm{ON}/I_\mathrm{OFF}$ ratio in tetrahedral amorphous carbon resistive \gls{MIM} devices\cite{2D-33}.
In addition, three-terminal memristors based on 2D materials have shown great promise due to the additional tunability and functionality provided though the additional gate terminal.
An example of three-terminal memristors is synaptic transistors, which utilize wide range of mechanisms, such as floating gate flash memory~\cite{2D-53} and gate-controlled charge trapping in gate dielectric~\cite{2D-28}.
On the other hand, \glspl{FeFET} utilize a ferroelectric layer in place of the gate dielectric.
As a result, nonvolatile states can be written to the device with gate control~\cite{2D-54}.
Finally, memtransistors operate similarly to its two terminal counterparts (memristors) with the exception that the resistance of the device is gate controlled.
In fact, several mechanisms governing resistive switching in memtransistors have been demonstrated, such as grain boundary migration~\cite{2D-31}, ferroelectric switching~\cite{2D-55}, and gate-controlled vdW heterojunctions~\cite{2D-56}.

Of course, with any of these technologies, due to the analog nature of computations, the idealized vector-matrix computation in \cref{fig:dpe-and-synaptic-layer:a} is often difficult to achieve.
Firstly, it may be challenging to set devices to the desired values of conductances $G_{i,j}$.
As an example, devices like memristors may get stuck in a certain conductance state~\cite{ZhUy2019} or even fail to electroform (i.e.\ become conductive)~\cite{YaMi2009}, experience \gls{RTN}~\cite{YaFi2019,PaKi2021} or programming variability~\cite{KiYa2016}, or have their conductance state drift over time~\cite{KiLu2013}.
Even more difficult to tackle are nonidealities that result in deviations from the linear (with respect to conductance and/or voltage) behavior, which \glspl{DPE} rely on; such nonidealities include \IV\ nonlinearity~\cite{SuLi2018,JoWa2022} and line resistance~\cite{Ch2013,SeRe2015,JoMe2020}.

There are multiple ways of utilizing \glspl{DPE} for the implementation of \glspl{ANN}.
The most obvious one has been alluded to earlier---neural network weights may be mapped onto crossbar conductances after they have been trained on digital computers.
However, it may also be possible to train \glspl{ANN} directly on crossbar arrays thus saving time, energy, and even preventing unnecessary greenhouse gas emissions.
That is attractive because training a large \gls{ANN} on a conventional digital architecture may emit as much \ce{CO2} as five cars throughout their lifetimes~\cite{StGa2019}.

\textit{Ex-situ} training is the most straightforward way of learning the weights of neural networks that are later implemented physically.
Such \glspl{ANN} can utilize a training process that is no different from the one used to train conventional networks.
Training on a digital computer is the simplest approach, but it obviously has disadvantages due to the mismatch between well-behaved conventional electronic systems and crossbar arrays consisting of analog devices.

If one does not take nonidealities into account, networks trained \textit{ex situ} may perform considerably worse on crossbar arrays, compared to their digital counterparts. 
For example, small number of achievable states, limited dynamic range, \gls{D2D} variability and \IV\ nonlinearities may all contribute to higher error rate~\cite{MeJo2019}.
In addition, system-level issues, including the aforementioned line resistance~\cite{LiHu2018,JoFr2020}, may disturb the distribution of currents and increase the error further.

This may be partly addressed by modifying \textit{ex-situ} training so that the nonidealities are considered \emph{before} deploying \glspl{ANN} onto \glspl{DPE}.
It is possible to model the behavior of analog devices, like memristors, and adjust the expected outputs of the hardware neural network accordingly.
Even for stochastic nonidealities, the nature of the stochasticity may inform the training process and make \glspl{ANN} more robust.
That is not unique to crossbar-based neural networks as noise can make even conventional \glspl{ANN} more robust~\cite{LiSi2019}.

There are multiple ways of taking nonidealities into account during training.
For example, the cost function (which quantifies how close \gls{ANN} outputs are to the expected ones) may be modified to incorporate the randomness associated with device behavior~\cite{ZhZh2020}.
Alternatively, network weights can be disturbed to represent nonidealities, like read and write noise~\cite{JoLe2020}.
Where the effects of nonidealities cannot be represented by injecting noise into the weights, their behavior can be redefined to reflect, for example, $I$-$V$ nonlinearities~\cite{JoWa2022}.

Although \textit{ex-situ} training can significantly improve the performance, it is important to consider that it relies on a number of assumptions.
If the modeling of nonidealities is inaccurate, that will be reflected in the training on a digital computer and may result in deviations from intended behavior when \glspl{ANN} are implemented physically.
However, this may be partly hedged against by including randomness in the modeling.
Randomness may represent the uncertainty in not only the device behavior, but also the designers' understanding of how the devices behave.
Therefore, it can improve the performance when the modeling is not perfectly accurate or even when different nonidealities manifest themselves~\cite{JoWa2022}.

Finally, one may employ \textit{in-situ} training, which can refer to either full or partial training directly on crossbar arrays.
Performing \gls{ANN} training on real devices can help networks adapt to specific instantiations of nonideal behavior---no two analog are the same, but \textit{in-situ}, unlike \textit{ex-situ}, training can take individual variations into account without the need to model the behavior.
\textit{In-situ} approach makes networks more robust to nonidealities, like faulty devices and \gls{D2D} variability~\cite{BuSh2015}.
One may even combine the two paradigms---conventional \textit{ex-situ} training can be used to produce \gls{ANN} weights, after which \textit{in-situ} retraining is used to recover from defects, like stuck devices~\cite{LiHu2017}.

Unfortunately, training networks \textit{in situ} is challenging.
Because conventional \gls{ML} methods rely on incremental adjustments of synaptic weights, analog devices may often be too unreliable for the task.
For example, the training process can be negatively affected by the asymmetry and nonlinearity of conductance changes~\cite{BuSh2015}, both of which are common in, for example, memristive devices.
Approaches for dealing with this include adjusting the fabrication process~\cite{WoMo2016,WuWu2018} and using digital electronic devices in conjunction with the analog ones~\cite{AmNa2018}.

\subsection{\Glsentrylongpl{SNN}}

Although \glspl{ANN} are loosely inspired by the brain, they are highly inefficient compared to biological systems.
This is due to the fact that there are fundamental differences between the two systems.
The adopted models of brain learning involve dynamic adjustment of synaptic strengths by the neuronal spiking activity.
In comparison, learning in \glspl{ANN} is based on gradient descent methods, which adjust weights in order to optimize an objective function.

There is a significant research interest in developing \glspl{SNN} as it is believed they could yield much better energy efficiency.
The fundamental difference is that in \glspl{SNN}, time is used directly to encode and process information---it is encoded in the time of arrival of binary events (``spikes'').
Two main functional units needed for the implementation of \glspl{SNN} are neurons and synapses. Neurons are typically implemented as simple leaky integrate-and-fire neurons, which are capable of integrating signals over time and producing spikes when a certain threshold is reached.
In terms of the synaptic functionalities, apart from adjustable strength, it is necessary to implement different local learning rules, such as spike-time-dependent plasticity, spike-rate-dependent plasticity, short-term plasticity, long-term potentiation, and long-term depression.

The energy efficiency argument relies on hopes of developing dedicated hardware platforms~\cite{ChDi2022} because current von~Neumann architectures are not best suited for the implementation of \gls{SNN} algorithms.
Although there exist many \gls{CMOS}-based implementations of \gls{SNN} hardware accelerators~\cite{BeGa2014,ScKl2017,LiPo2008,MoQi2017,WaTh2017,DiCo2015,FuGa2014,DaSr2018,PeDe2019,ChKu2019}, these systems are still lacking in terms of the energy efficiency of biological counterparts.
It is believed that emergent technologies will be able to directly implement critical functionalities using voltages and currents much lower than \gls{CMOS} equivalents~\cite{SpSe2020}.

Memristive technology has been used to implement multiple elements of the \gls{SNN} paradigm.
Synaptic functionalities were implemented by incorporating temporal plasticity as well as particular local learning rules~\cite{JoCh2010,WaJo2017,ZaMe2018}.
\Gls{PCM} memristors\cite{TuPa2016}, \glspl{ReRAM}~\cite{DiNa2016,GuSe2016}, and Mott-based memristors~\cite{PiMe2013,KuWi2020} have all been used for emulating neuronal activity.
For more details and a much more comprehensive overview of using memristors for \glspl{SNN}, we refer readers to Ref.~\cite{SpSe2020}.

\begin{figure*}[h]
  \centering
  \includegraphics{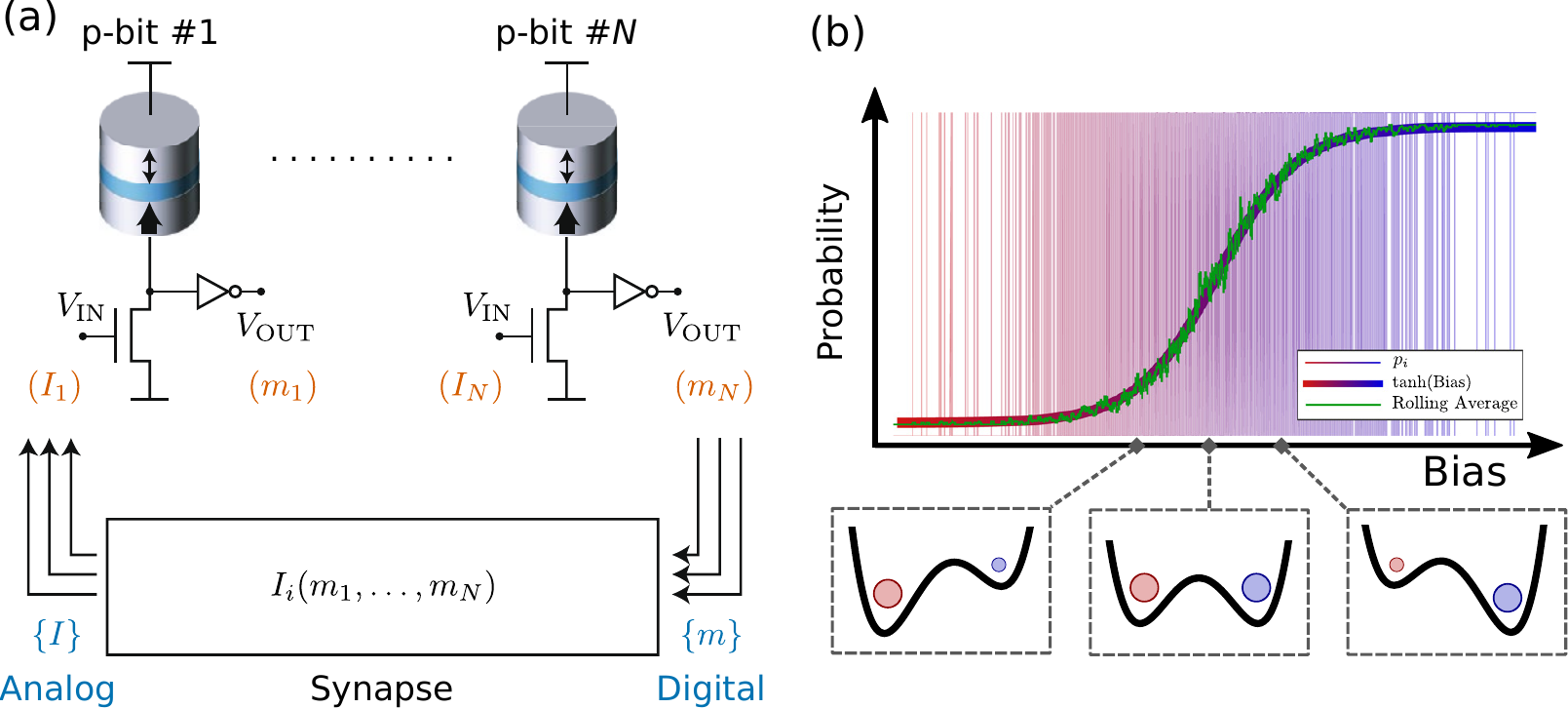}
  {\phantomsubcaption\label{fig:spiking:a}}
  {\phantomsubcaption\label{fig:spiking:b}}
  \mycaption{%
    Spintronic approaches to \glsentrylongpl{SNN}%
  }{%
    (a)~Schematic representation of p-bit computing scheme.
    Superparamagnetic tunnel junctions offer extremely low energy barriers, which can be exploited to solve complex problems.
    The analog input voltage to some junction, $I$, can cause a nonlinear response to the digital output voltages as shown in (b), and form random fluctuations analogous to 0's and 1's of a stochastic neuron at room temperature. 
    Adapted from~\cite{grollier_NatureElec2020}.
    (b)~The control of bias voltages changes the relative energies of two states.
    Adapted from~\cite{Camsari_APR2019}.
  }\label{fig:spiking}
\end{figure*}

Spintronic devices, too, may be used for physical implementations of \glspl{SNN}.
The oscillatory behavior of biological neurons can be emulated using \glspl{STNO}~\cite{Yogendra_IEEE2015, Yogendra_IEEE2016}; the required power may be achieved when assisted by a microwatt nanosecond laser pulse~\cite{Farkhani_IEEE2019}.
When the system is configured towards the limit of super-paramagnetism, the random spiking of biological neurons can be emulated to perform population coding and probabilistic computing~\cite{mizrahi_NatComs2018, Camsari_APR2019}.
\Cref{fig:spiking:a} shows a schematic of probabilistic computing with probabilistic-bits (p-bits), where the structural design of the \glspl{MTJ} benefits from the low-energy-barrier of the superparamagnetic tunnel junctions.
The analog input voltage, $I$, to some junction can cause a nonlinear response to the digital output voltage, $m$, (\cref{fig:spiking:b}) and form random fluctuations analogous to 0's and 1's of a stochastic neuron at room temperature.
Nevertheless, other systems such as memristors or nano-arrays or exploiting nonlinear dynamics in variant forms of magnetic spin textures like domain walls or skyrmions can also be engineered to facilitate such properties~\cite{huang_IOPNano2017, li_IOPNano2017, grollier_IEEE2016, chen_Nanoscale2018, chen_IEEE2018}, demonstrating the potential of spintronic devices as artificial neuromorphic components.

Photonic circuits represent another possible approach to neuromorphic computing and \glspl{SNN} in particular~\cite{2D-35}.
For example, black phosphorus has been used to emulate excitatory and inhibitory action potentials by using oxidation-related defects~\cite{2D-36}.
Also, \ce{WSe2}/\glsentryshort{hBN} heterostructures have been used as $7$-bit non-volatile optoelectronic memories~\cite{2D-37} and for colored and mixed color pattern recognition~\cite{2D-38}.
Further, the developments in the field of optoelectronic memristive devices~\cite{XuCi2020} could provide further flexibility and extended functionality, such as in-sensory computing~\cite{WaMe2021}.
In many cases, the operation of these devices requires both electronic and optical stimulation~\cite{MeGe2017}; however, fully optically operable memristors can be realized~\cite{HuYa2021} with favorable properties for neuromorphic computing.

\subsection{Reservoir computing}

In addition to the aforementioned fully connected \glspl{ANN}, there also exist \glspl{RNN}.
These networks contain recurrent connections and can be incredibly useful when dealing with time series data~\cite{CoMa1994}.
However, \glspl{RNN} can suffer from vanishing and exploding gradients, which makes their training especially difficult~\cite{PaMi2013}.

\begin{figure*}[h!]
  \centering
  \includegraphics{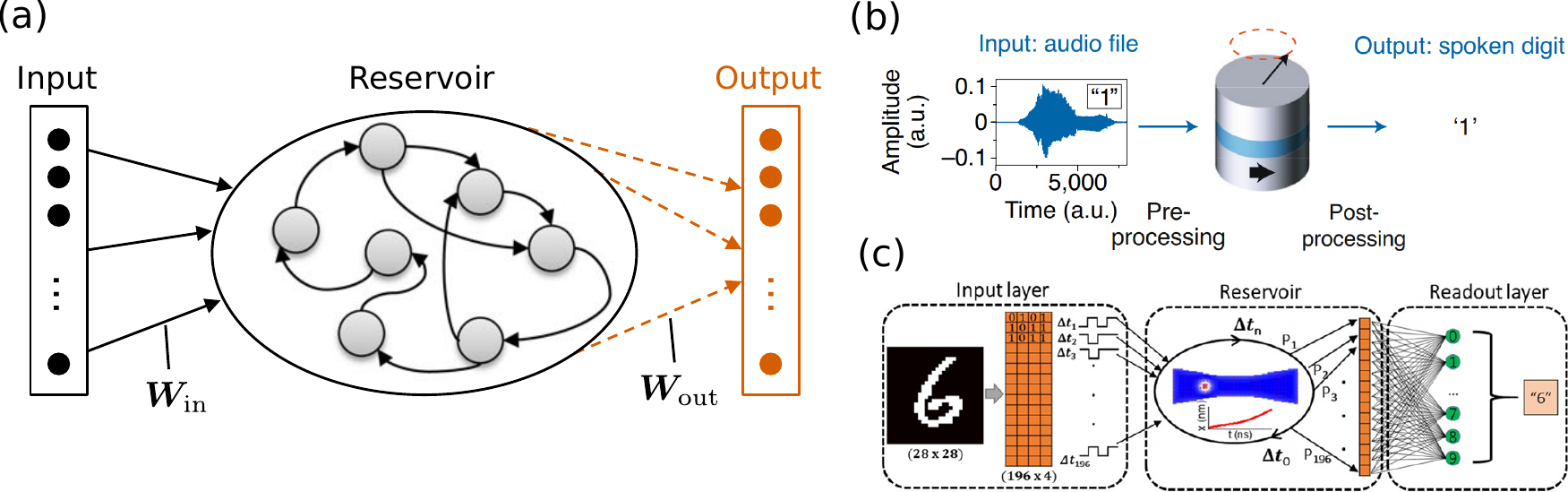}
  {\phantomsubcaption\label{fig:reservoir-computing:a}}
  {\phantomsubcaption\label{fig:reservoir-computing:b}}
  {\phantomsubcaption\label{fig:reservoir-computing:c}}
  \mycaption{%
    Operating principles and example implementations of reservoir computing%
  }{%
    (a)~Inputs and the interconnected nonlinear units of the reservoir are connected through a set of weights, $\matr{W}_\mathrm{in}$.
    Those reservoir nodes and the outputs are connected through another set of weights, $\matr{W}_\mathrm{out}$; during training, only $\matr{W}_\mathrm{out}$ need to be learned.
    Adapted from~\cite{ZhTa2021}.
    (b)~Experimental implementation of physical reservoir computing using \glsentrylong{STNO} for spoken digit recognition.
    Adapted from~\cite{grollier_NatureElec2020}.
    (c)~Numerical demonstration of physical reservoir computing scheme using skyrmion positions for classification of hand-written digits.
    Adapted from~\cite{jiang_APL2019}.
  }\label{fig:reservoir-computing}
\end{figure*}

Given the challenges of \glspl{RNN}, reservoir computing has been suggested as an alternative~\cite{Ja2001}.
It relies on systems that exhibit rich dynamic behaviors to do the computations ``for free.''
Like activation functions in conventional \glspl{ANN} may introduce nonlinearities, physical ``reservoirs'', which are complex, nonlinear, and have short-term memory properties, are able to map inputs to the nonlinear dynamics of a high-dimensional system.
This enables to perform training only on the last synaptic---and usually linear---layer.
The principles behind reservoir computing are visualized in \cref{fig:reservoir-computing:a}.

Many kinds of memristors hold promise as potential mediums of reservoir computing.
One of the factors enabling this is the fact that many memristors exhibit short-term memory properties.
In the case of some memristors, repeatedly applying voltage pulses may gradually increase the response, while the absence of the pulses will make the devices decay toward their original resting state~\cite{DuCa2017}.
Additionally, nonlinear \IV\ characteristics of memristive devices can be incredibly useful for reservoir computing applications~\cite{ZhTa2021}.

One may also use spintronic devices in reservoir computing applications.
\Cref{fig:reservoir-computing:b} shows an experimental demonstration of using a single \gls{STNO} facilitated with \pgls{MTJ} as a reservoir.
It exploited time multiplexing to emulate up to \num{400} neurons by tuning the state of each neuron at periodic intervals.
The relationships between the input current and the oscillation frequency can bring a nonlinear response, and the motion of spins in the free layer showed history dependence as a response to the amplitudes of analog audio signals.
Another example has been demonstrated by exploiting spinwaves in a three-dimensional space using small-sized metal electrodes to apply and detect the input and output voltages (currents)~\cite{nakane_IEEE2018}. The system was configured as a stacked device consisting a thin yttrium iron garnet layer between the conductive substrate and magneto-electric coupling layer. The nonlinear effects and the history-dependent motion of the spinwaves were achieved by controlling the stability of the precession of the spins by reducing the applied bias DC magnetic field, allowing the device to satisfy the reservoir computation criteria.
Yet another proposed medium for reservoir computing has been magnetic skyrmions due to their stability and controllable history-dependent nonlinear effects.
In an example design in Ref.~\cite{jiang_APL2019} (shown in \cref{fig:reservoir-computing:c}), handwritten digits were converted into an input sequence of current pulses, which was fed into a magnetic skyrmion memristor.
The nonlinear relationship between the positions of the magnetic skyrmions allowed the system to be configured as a physical reservoir.
In addition to this approach, a wide range of different systems have been proposed and investigated, including the manipulation of skyrmion fabrics, skyrmion position, and interaction of multiple skyrmions~\cite{bourianoff_AIP2018, prychynenko_PRA2018, pinna_PRA2020}.

\section{Outlook and Conclusion}

Here we discuss the basics of three emerging nanoscale technologies with great potential to improve and extend the infrastructure of compute hardware.
One plausible scenario that addresses the growing diversity and complexity of computational problems includes a synergy between more conventional, digital systems and new paradigms of computing hardware.
General-purpose computing will likely remain best implemented on digital systems, which use Boolean logic and higher precision computing.
However, some applications, like \glspl{ANN}, which are currently realized on these digital systems, could benefit from speed and power efficiency that neuromorphic hardware accelerators offer.
Further, computing approaches like \glspl{SNN}, which are even less fit for conventional computers, could be implemented using devices that exhibit more exotic behaviors, such as synaptic plasticity or neuronal spiking.
Finally, there are paradigms of computing that are feasible or possible only with devices that exhibit certain physical behavior; an example of this is reservoir computing.
Memristive, spintronic, and 2D-materials-based devices will likely play a role in both the improvement of digital hardware and the adoption of more novel approaches.

Many systems would benefit from fast low-power memristive hardware, but, at the same time, some are constrained by additional requirements.  
For example, memristive \glspl{ANN} could in theory be used by autonomous driving companies; however, these companies often utilize driving data to improve their \gls{ML} models and deploy the updated models continuously~\cite{Co2022,Te2022}.
Even if \glspl{ANN} are trained \textit{ex situ} and identical versions are deployed onto memristive systems, each physical instantiation will be at least slightly different.
This could affect not only the behavior of individual vehicles, but also the \gls{ML} pipeline, i.e.\ data that are collected and then used to improve the models~\cite{ZhYu2020} that are deployed to \emph{all} cars.
In general, we can identify multiple challenges of memristive systems that need to be addressed before wide-scale deployment in the real world:
\begin{itemize}
  \item non-identical behavior of identically designed systems~\cite{ChZh2011}
  \item stochasticity, including possibly changing behavior over time~\cite{TiLi2019,LiLo2021}
  \item difficulty of reprogramming once deployed in the real world
  \item difficulty of identifying hardware faults~\cite{DoLi2016,ChCh2021a}
\end{itemize}

Where safety and behavior reproducibility are key, special attention currently needs to be paid to the treatment of device stochasticity, variability and reliability.
This is especially true when memristors are used unconventionally (i.e.\ not for digital nonvolatile memory, but as analog memory and neuromorphic computational primitives).
Similarly, applications where hardware needs to be constantly reconfigured (e.g.\ updating \gls{ML} models in autonomous vehicles) would be challenging---even in controlled environments, programming memristive devices remains difficult~\cite{KiYa2016,JaCh2021}.
In addition, cycling endurance might need to be improved to match the endurance of volatile memory (e.g.\ $10^{16}$ cycles in \gls{SRAM}).

We believe that memristors can be the most useful where computing needs to be fast, low-power and/or local (i.e.\ not in the cloud).
The last possibility flows from the first two---data-intensive applications like \glspl{ANN} consume a lot of power, thus the computing often takes place remotely; however, memristive technologies---due to their speed and power efficiency---can enable to perform the computations locally~\cite{PhTr2019,KrJa2019}.
We therefore believe that these devices are very well suited for applications like the \gls{IoT} where potential violations of privacy remain a significant issue~\cite{LeAh2021}.
Memristive implementations of data-intensive tasks would not only eliminate the need to send data to the server, but also ensure low-power operation and high speed.

Spintronics is another promising approach that can advance the state-of-the-art in multiple paradigms of computing. Spintronic memory and logic circuits are expected to open a novel route to manipulate information more efficiently and their prototypes have been actively proposed~\cite{Guo_IEEE2021,Finocchio_JMMM2020,Dieny_NatElec2020,Shao_TMAG2021}.
In the coming decade, we predict an increased dominance of hybrid \gls{CMOS}-spintronic computing architectures based on \gls{MRAM} techniques such as \gls{STT}, \gls{SOT} and \gls{VCMA}.
Moreover, the desired progress in speed, energy and scaling will also require the use of advanced materials such as antiferromagnets~\cite{Jungwirth_Nnano2016}, 2D materials~\cite{Lin_NElec2019,Kurebayashi_NRevPys2022}, topological insulators~\cite{Pesin_NMater2012}.
Spintronic devices are also being employed in a new class of computer architecture such as \gls{ASL}~\cite{Kim2014} and \gls{LIM}~\cite{Matsunaga2009}.
\Gls{LIM} structures are hybrid in nature, combining contemporary spintronics components, such as \glspl{MTJ}, with current \gls{CMOS} devices.
Advancement in fabrication technology (e.g.\ 3D back-end process) enabled the growth of \glspl{MTJ} on the silicon layer without compromising the functionality of the circuit~\cite{Tehrani1999}.
Circuits developed using \gls{LIM} hold advantages over the conventional \gls{CMOS} technologies due to their lower power dissipation, non-volatility, high density, fast reading capability, infinite endurance and 3D fabrication adaptability~\cite{Verma2016}.  

The properties of spintronic devices (e.g.\ high-speed dynamics of GHz to potentially THz ranges, nonvolatilty, plasticity and nonlinearity) offer ample room for accessing numerous building blocks that can mimic the key features of biological synapses and neurons~\cite{Romera_Nature2018, torrejon_Nature2017, song_NatureElect2020, huang_IOPNano2017, li_IOPNano2017, grollier_IEEE2016, chen_Nanoscale2018, chen_IEEE2018}.
In spintronic devices, the processing/transfer of information can be achieved via spin currents, spin waves, microwave signals, or magnetic spin textures such as domain walls and skyrmions.
Such properties can potentially find their unique positions in the electronics market by offering a more compact and energy-efficient approaches, exploiting the spin degree of freedom.

While proof-of-concept spintronics-based neuromorhic computing implementations have been demonstrated~\cite{Romera_Nature2018, torrejon_Nature2017, grollier_NatureElec2020,Romera2022}, there remain a number of key challenges.
Although many creative and exciting ideas have been proposed, it is important to consider the viability of mass production and scalability when it comes to spintronics-based neuromorphic computing.
Likewise, traditional algorithms used on \gls{CMOS} technology require enhanced tuning to harness the maximum potential of such spintronic neuromorphic chips.
Similar to von~Neumann architecture for conventional computing, a dedicated architecture is a prerequisite for wide-scale implementation of neuromorphic computing~\cite{Christensen_2022}.
Furthermore, additional research is required to increase the capability of the proposed devices.
For example, enhancing the coupling efficiency between the \gls{MTJ} layers and the relatively low ratio of maximum to minimum resistance of the existing devices~\cite{grollier_NatureElec2020}.

2D materials are yet another key enabler for future computing technologies.
Taken individually, or in combination to form heterostructures with tailored properties, they offer an unprecedented playground for both conventional and emerging forms of computing.
However, there are a number of challenges to overcome before their full potential is realized.

The first is the doping because the ion implantation processes commonly used in semiconductor industry are not applicable to 2D materials due to their atomic-thickness~\cite{2D-46}.
Instead of replacing atoms in the crystal lattice (as in substitutional doping used for 3D semiconductors), doping in 2D materials is normally achieved either by physisorption, covalent bonding of impurities (chemical doping), or proximity with compounds, which modifies the dielectric environment and leads to local gating effect (sometimes referred to as solid-state doping)~\cite{2D-47}.
Unfortunately, to date, none of these methods fully satisfy the stringent requirement of ultra-scaled devices and more research effort should be devoted to identifying an industry-compatible, precise, stable and reproducible doping method.

The second challenge to overcome is related to the deposition of high-$\kappa$ dielectrics.
Indeed, the lack of dangling bonds in 2D materials' surfaces complicates the growth of thin, uniform insulating layers by atomic layer deposition and, often, ``seed'' layers are required to facilitate the growth.
Dielectrics are not only important for the functionality of devices (e.g.\ as gate dielectric in \glspl{MOSFET}) but also to encapsulate 2D materials, as their properties are often significantly degraded by substrate, contamination, roughness, and charged impurities.
A promising alternative is represented by 2D dielectrics, which form atomically-sharp interfaces with other 2D materials.
\Gls{hBN} is by far the most explored 2D dielectric, which enabled experimental investigation of transport phenomena and proof-of-concept devices~\cite{2D-48,2D-49}.
However, low dielectric constant (\num{\sim 3}) and difficulty in scalable production of multi-layer \gls{hBN} limits its applicability in high-performance computing.
A more promising option is represented by the possibility of oxidizing hafnium and zirconium-based multilayer \glspl{TMD} to form high-$\kappa$ dielectrics \ce{HfO2} and \ce{ZrO2}~\cite{2D-50,2D-51}.
This approach is of particular interest as it is the equivalent to the oxidation of silicon and results in almost-perfect interfaces between the pristine semiconducting part and the oxidized surface.

The third challenge is represented by contacts.
Contact resistance is usually high and cannot be reduced by ion implantation as in 3D semiconductors.
Moreover, due to the Schottky junction formed when depositing metals on 2D semiconductor, contact resistance is also modified by applied gate voltage, introducing additional delays and complicating the analysis of devices~\cite{2D-46}.
Theoretical and experimental effort should be devoted towards this essential but often disregarded aspect of computing.
Finally, scalable production of 2D materials should be optimized, in particular for what concerns reproducibility and control over defects and contaminations.
\Gls{CVD} growth has made impressive progress in the last ten years, however some fundamental challenges remain, such as the lack of an industrially-scalable, clean transfer of graphene.
Our view is that 2D materials do not represent a replacement, but rather a complement to current bulk semiconductor technology.
The relative ease of integrability of such materials into established semiconductor production lines will indeed be the key for a synergy between the two technologies and enable new, high performing computing.

Memristors, spintronics and 2D materials are rapidly developing and changing fields.
New developments span materials, devices, circuit/system design and algorithmic approaches.
This perspective article provides a basic introduction to central ideas, explores potential advantages over conventional \gls{CMOS} technologies, and lists some pressing challenges that still need to be addressed.
Memristors, spintronics and 2D-based electronics are among the most promising candidates for supporting future computing systems.
There is a strong possibility they will co-exist and complement other emerging technologies and approaches, as well as conventional electronics systems.

\section*{Acknowledgements}

A.\,M.\ acknowledges funding from the Royal Academy of Engineering under the Research Fellowship scheme, A.\,J.\,K.\ acknowledges funding from the Engineering and Physical Sciences Research Council (EP/P013503/1) and the Leverhulme Trust (RPG-2016-135), D.\,J.\ acknowledges studentship funding from the Engineering and Physical Sciences Research Council (ref.\ 2094654), A.\,A.\ acknowledges funding from the Saudi Ministry of Education.

\section*{Conflict of Interest}

A.\,M.\ and A.\,J.\,K.\ are co-founders of Intrinsic, a company developing memristor technology.

\printbibliography

\end{document}